\IfFileExists{/dev/null}{\immediate\write18{sh makeLinks.sh}}{\immediate\write18{makeLinks.bat}}

\documentclass[lettersize,journal]{IEEEtran}

\makeatletter
\def\subsubsection{%
	\@startsection
	{subsubsection}                 
	{3}                             
	{\z@}                           
	{2.5ex plus 1.5ex minus 1.5ex}  
	{1ex plus .5ex minus 0ex}     
	{\normalfont\normalsize\itshape}
}
\makeatother

\usepackage[none]{hyphenat}
\usepackage[utf8]{inputenc}
\usepackage[T1]{fontenc}
\usepackage{textcomp}
\usepackage{scalefnt}
\usepackage{amsmath,amsfonts,amssymb,amsbsy,amstext}
\usepackage[normalem]{ulem}
\useunder{\uline}{\ul}{}

\usepackage{mathtools}          
\usepackage{cmap}               
\usepackage[dvipsnames,svgnames,x11names]{xcolor}   
\usepackage{varwidth}

\let\originalleft\left
\let\originalright\right
\renewcommand{\left}{\mathopen{}\mathclose\bgroup\originalleft}
\renewcommand{\right}{\aftergroup\egroup\originalright}

\usepackage{graphicx}
\graphicspath{{.}{.\figs}}
\usepackage{placeins} 

\usepackage{multirow}
\usepackage{tabularx}
\newcolumntype{C}{>{\centering\arraybackslash}X}
\newcolumntype{R}{>{\flushright\arraybackslash}X}
\newcolumntype{L}{>{\flushleft\arraybackslash}X}
\newcolumntype{P}{>{\centering\arraybackslash} p{0.5\linewidth}}
\usepackage{booktabs}

\usepackage[%
style=ieee,
backend=biber, 
]{biblatex}

\defbibheading{bibliography}[\refname]{\section*{#1}}

\makeatletter
\g@addto@macro{\UrlBreaks}{\UrlOrds}
\makeatother

\usepackage{acro}[=v2]

\NewDocumentCommand{\acro}{m o m o}
{%
	\IfValueTF{#2}{%
		\IfValueTF{#4}{%
			\DeclareAcronym{#1}{short={#2},long={#3},#4}
		}{%
			\DeclareAcronym{#1}{short={#2},long={#3}}
		}
	}{%
		\IfValueTF{#4}{%
			\DeclareAcronym{#1}{short={#1},long={#3},#4}
		}{%
			\DeclareAcronym{#1}{short={#1},long={#3}}
		}
	}
}

\acro{2D}{two-dimensional}
\acro{3D}{three-dimensional}
\acro{4D}{four-dimensional}
\acro{6D}{six-dimensional}
\acro{1G}{first generation}
\acro{2G}{second generation}
\acro{3G}{third generation}
\acro{4G}{fourth generation}
\acro{5G}{fifth generation}
\acro{5GC}{5G core network}
\acro{5G-StoRM}{5G stochastic radio channel for dual mobility}
\acro{6G}{sixth generation}
\acro{3GPP}{3rd generation partnership project}
\acro{3GPP2}{3rd generation partnership project 2}
\acro{A2A}{air-to-air}
\acro{A2G}{air-to-ground}
\acro{AA}{antenna array}
\acro{AC}{admission control}
\acro{ACK}{acknowledge}
\acro{AD}{attack-decay}
\acro{AE}{antenna element}
\acro{AF}{amplify and forward}
\acro{ABS}{almost blank subframe}
\acro{ACF}{autocorrelation function}
\acro{ADSL}{asymmetric digital subscriber line}
\acro{AHW}{alternate hop-and-wait}
\acro{AI}{artificial intelligence}
\acro{AMC}{adaptive modulation and coding}
\acro{AoA}{angle of arrival}
\acro{AoD}{angle of departure}
\acro{AP}{access point}
\acro{APA}{adaptive power allocation}
\acro{API}{application protocol interface}
\acro{ARQ}{automatic repeat request}
\acro{AT}{averaged time}
\acro{AR}{autoregressive}
\acro{ARMA}{autoregressive moving average}
\acro{ASE}{average squared error}
\acro{ASC}{adaptive satisfaction control}
\acro{ATES}{adaptive throughput-based efficiency-satisfaction trade-off}
\acro{AWGN}{additive white gaussian noise}
\acro{B5G}{beyond 5G}
\acro{BAP}{backhaul adaptation protocol}
\acro{BB}{branch and bound}
\acro{BC}{branch and cut}
\acro{BD}{block diagonalization}
\acro{BER}{bit error rate}
\acro{BF}{best fit}
\acro{BL}{bit loading}
\acro{BLER}{block error rate}
\acro{BLPC-1}{bit loading with power constraint in hop 1}
\acro{BLPC-2}{bit loading with power constraint in hop 2}
\acro{BPC}{binary power control}
\acro{BPSK}{binary phase-shift keying}
\acro{BRA}{balanced random allocation}
\acro{BS}{base station}
\acro{BSP}{\acs*{BS} placement}
\acro{BSR}{buffer status report}
\acro{BoI}{band of interest}
\acro{C-link}{control link}
\acro{CAP}{combinatorial allocation problem}
\acro{CAPEX}{capital expenditure}
\acro{CB}{contextual bandit}
\acro{CBF}{coordinated beamforming}
\acro{CBR}{constant bit rate}
\acro{CBS}{class based scheduling}
\acro{CC}{congestion control}
\acro{CCL}{common cell list}
\acro{CDF}{cumulative distribution function}
\acro{CDL}{clustered delay line}
\acro{CDMA}{code-division multiple access}
\acro{CIR}{channel impulse response}
\acro{CH}{channel hardening}
\acro{CHO}{conditional handover}
\acro{C-RAN}{cloud-based radio access network}
\acro{CL}{closed loop}
\acro{CLT}{central limit theorem}
\acro{CLPC}{closed loop power control} 
\acro{CN}{core network}    
\acro{CNN}{convolutional neural network}   
\acro{CNR}{channel-to-noise ratio}
\acro{CP}{control plane}
\acro{CPA}{cellular protection algorithm}
\acro{CPICH}{common pilot channel}
\acro{CoMP}{coordinated multi-point}
\acro{CQI}{channel quality indicator}
\acro{CRE}{cell range expansion}
\acro{CRM}{constrained rate maximization}
\acro{CRN}{cognitive radio network}
\acro{C-RNTI}{cell radio network temporary identifier}
\acro{CRRM}{centralized/common radio resource management}
\acro{CRS}{cell-specific reference signal}
\acro{CS}{coordinated scheduling}
\acro{CSI}{channel state information}
\acro{CSI-RS}{channel state information reference signal}
\acro{CTS}{clear to send}
\acro{CU}{centralized unit}
\acro{CUE}{cellular user equipment}
\acro{CWND}{congestion window size}
\acro{D2D}{device-to-device}
\acro{D2N}{drone-to-network}
\acro{DAA}{drone antenna array}
\acro{DAG}{directed acyclic graph}
\acro{DBS}{drone mounted base station}
\acro{DC}{dual connectivity}
\acro{DCA}{dynamic channel allocation}
\acro{DE}{differential evolution}
\acro{DF}{decode and forward}
\acro{DFT}{discrete fourier transform}
\acro{DIST}{distance}
\acro{DL}{downlink}
\acro{DMA}{double moving average}
\acro{DMRS}{demodulation reference signal}
\acro{D2DM}{\acs*{D2D} mode}
\acro{DMS}{\acs*{D2D} mode selection}
\acro{DP}{data plane}
\acro{DPC}{dirty paper coding}
\acro{DRA}{dynamic resource assignment}
\acro{DRX}{discontinuous reception}
\acro{DSA}{dynamic spectrum access}
\acro{DSM}{delay-based satisfaction maximization}
\acro{DU}{distributed unit}
\acro{E2E}{end-to-end}
\acro{ECC}{electronic communications committee}
\acro{EDF}{earliest deadline first}
\acro{EE}{energy efficiency}
\acro{EFLC}{error feedback based load control}
\acro{EI}{efficiency indicator}
\acro{e-ICIC}{enhanced inter-cell interference coordination}
\acro{eMBB}{enhanced mobile broadband}
\acro{ETSI}{European Telecommunications Standards Institute}
\acro{EM}{electromagnetic}
\acro{eNB}{evolved node B}
\acro{EXP}{exponential}
\acro{EPA}{equal power allocation}
\acro{EPC}{evolved packet core}
\acro{EPS}{evolved packet system}
\acro{E-UTRAN}{evolved universal terrestrial radio access network}
\acro{ES}{exhaustive search}
\acro{FCP}{fundamental counting principle}
\acro{FCA}{flow control algorithm}
\acro{FD}{full duplex}
\acro{FDD}{frequency division duplex}
\acro{FDM}{frequency division multiplexing}
\acro{FDMA}{frequency division multiple access}
\acro{FER}{frame erasure rate}
\acro{FIFO}{first in first out}
\acro{FoV}{field-of-view}
\acro{FF}{fast fading}
\acro{FL}{federated learning}
\acro{FR}{frequency range}
\acro{FS}{fast switching}
\acro{FSB}{fixed switched beamforming}
\acro{FST}{fixed \acs*{SNR} target}
\acro{FT}{fourier transform}
\acro{FTP}{file transfer protocol}
\acro{Fwd}{forwarding}
\acro{GA}{genetic algorithm}
\acro{GAP}{generalized assignment problem}
\acro{GAP-MQ}{generalized assignment problem with minimum quantities}
\acro{GATES}{generalized adaptive throughput-based efficiency-satisfaction trade-off}
\acro{GBR}{guaranteed bit rate}
\acro{GLR}{gain to leakage ratio}
\acro{gNB}{gNodeB}
\acro{GOS}{generated orthogonal sequence}
\acro{GP}{gaussian process}
\acro{GPL}{GNU general public license}
\acro{GPS}{global positioning system}
\acro{GRP}{grouping}
\acro{GR}{group report}
\acro{GSM}{global system for mobile communications}
\acro{GTEL}{wireless telecommunications research group}
\acro{HAP}{high altitude platform}
\acro{HARQ}{hybrid automatic repeat request}
\acro{HBF}{hybrid beamforming}
\acro{HCPP}{hardcore point process}
\acro{HetNet}{heterogeneous network}
\acro{HD}{half duplex}
\acro{HF}{high-frequency}
\acro{HH}{hughes-hartogs}
\acro{HMS}{harmonic mode selection}
\acro{HO}{handover}
\acro{HOL}{head of line}
\acro{HPBW}{half power beamwidth}
\acro{HSDPA}{high speed downlink packet access}
\acro{HSR}{high-speed railway}
\acro{HSPA}{high speed packet access}
\acro{HTTP}{hypertext transfer protocol}
\acro{IAB}{integrated access and backhaul}
\acro{ICMP}{internet control message protocol} 
\acro{ICI}{intercell interference}
\acro{ICIC}{inter-cell interference coordination}
\acro{ID}{identification}
\acro{i.i.d.}{independent and identically distributed}
\acro{IETF}{internet engineering task force}
\acro{IPC}{individual power constraint}
\acro{ISAC}{integrated sensing and communication}
\acro{UID}{unique identification}
\acro{IID}{independent and identically distributed}
\acro{IIR}{infinite impulse response}
\acro{ILP}{integer linear problem}
\acro{IMT}{international mobile telecommunications}
\acro{INV}{inverted norm-based grouping} 
\acro{IoT}{internet of things}
\acro{IP}{internet protocol}
\acro{IPv6}{internet protocol version 6}
\acro{IRA}{integrated resource allocation}
\acro{IRS}{intelligent reflecting surface}
\acro{ISD}{inter-site distance}
\acro{ISI}{inter symbol interference}
\acro{ISM}{industrial, scientific and medical}
\acro{ITU}{international telecommunication union}
\acro{JOAS}{joint opportunistic assignment and scheduling}
\acro{JOS}{joint opportunistic scheduling}
\acro{JP}{joint processing}
\acro{JRAPAP}{joint rb assignment and power allocation problem}
\acro{JS}{jump-stay}
\acro{JSM}{joint satisfaction maximization}
\acro{KKT}{karush-kuhn-tucker}
\acro{KPI}{key performance indicator}
\acro{LAC}{link admission control}
\acro{LA}{link adaptation}
\acro{LAP}{low altitude platform}
\acro{LBS}{location based service}
\acro{LC}{load control}
\acro{LEB}{load-efficiency-balance}
\acro{LOS}{line of sight}
\acro{LP}{linear programming}
\acro{LSTM}{long short-term memory}
\acro{LTE}{long term evolution}
\acro{LTE-A}{\acs*{LTE}-advanced}
\acro{LTE-Advanced}{\ac{LTE-A}}
\acro{LTE-R}{\ac{LTE} railway}
\acro{LSP}{large scale parameter}
\acro{MADRDPG}{multi-agent deep recurrent deterministic policy gradient}
\acro{MeNB}{master \acs*{ENB}}
\acro{M2M}{machine-to-machine}
\acro{MAC}{medium access control}
\acro{MANET}{mobile ad hoc network}
\acro{MEDS}{method of exact doppler spread}
\acro{MC}{modular clock}
\acro{MCP}{minimal cost power}
\acro{MCS}{modulation and coding scheme}
\acro{MDB}{measured delay based}
\acro{MDI}{minimum \acs*{D2D} interference}
\acro{MDSM}{modified delay-based satisfaction maximization}
\acro{MDU}{max-delay-utility}
\acro{METIS}{mobile and wireless communications enablers for the twenty-twenty information society \acs*{5G}}
\acro{MF}{matched filter}
\acro{MFAP}{mobile femtocell access point}
\acro{MG}{maximum gain}
\acro{MH}{multi-hop}
\acro{mIAB}{mobile \ac{IAB}}
\acro{MILP}{mixed integer linear programming}
\acro{MIMO}{multiple input multiple output}
\acro{MINLP}{mixed integer nonlinear programming}
\acro{MIP}{mixed integer programming}
\acro{MISO}{multiple input single output}
\acro{MIT}{mobility interruption time}
\acro{ML}{machine learning}
\acro{MLWDF}{modified largest weighted delay first}
\acro{MME}{mobility management entity}
\acro{MMF}{max-min fairness}
\acro{mmMAGIC}{millimetre-wave based mobile radio access network for fifth generation integrated communications}
\acro{MMRP}{maximizing the minimal residual power}
\acro{MMRP-LB}{maximizing the minimal residual power with lower bound}
\acro{MMSE}{minimum mean square error}
\acro{mMTC}{massive machine-type communications}
\acro{mmWave}{millimeter wave}
\acro{MN}{master node}
\acro{MOS}{mean opinion score}
\acro{MPF}{multicarrier proportional fair}
\acro{MPRP}{maximization of the product of the residual powers}
\acro{MRA}{maximum rate allocation}
\acro{MR}{maximum rate}
\acro{MRC}{maximum ratio combining}
\acro{MRN}{mobile relay node}
\acro{MRT}{maximum ratio transmission}
\acro{MRUS}{maximum rate with user satisfaction}
\acro{MS}{mode selection}
\acro{MSE}{mean squared error}
\acro{MCG}{master cell group} 
\acro{MSI}{multi-stream interference}
\acro{MT}{mobile termination}
\acro{MTC}{machine-type communication}
\acro{IMS}{\acs*{IP} multimedia subsystem}
\acro{MTSI}{multimedia telephony services over \acs*{IMS}}
\acro{MTSM}{modified throughput-based satisfaction maximization}
\acro{MU-MIMO}{multi-user multiple input multiple output}
\acro{MU}{multi-user}
\acro{Multi-CUT}{multi-cell and multi-user and multi-tier}
\acro{NAS}{non-access stratum}
\acro{NACK}{no acknowledge}
\acro{NB}{node B}
\acro{nBS}{neighbor base station}
\acro{NCL}{neighbor cell list}
\acro{NCR}{network-controlled repeater}
\acro{NG}{next generation}
\acro{NGC}{next generation core network}
\acro{NLP}{nonlinear programming}
\acro{NLOS}{non-line of sight}
\acro{NMSE}{normalized mean square error}
\acro{NN}{neural network}
\acro{NOMA}{non-orthogonal multiple access}
\acro{NORM}{normalized projection-based grouping}
\acro{NP}{non-polynomial time}
\acro{NR}{new radio}
\acro{NRT}{non-real time}
\acro{NSA}{non-stand-alone}
\acro{NSPS}{national security and public safety services}
\acro{OAM}{operation and maintenance}
\acro{OBF}{opportunistic beamforming}
\acro{OFDMA}{orthogonal frequency division multiple access}
\acro{OFDM}{orthogonal frequency division multiplexing}
\acro{OFPC}{open loop with fractional path loss compensation}
\acro{O2I}{outdoor-to-indoor}
\acro{OL}{open loop}
\acro{OLPC}{open-loop power control}
\acro{OL-PC}{open-loop power control}
\acro{OMA}{orthogonal multiple access}
\acro{OPEX}{operational expenditure}
\acro{ORA}{orthogonal resource allocation}
\acro{ORB}{orthogonal random beamforming}
\acro{OSM}{open street map}
\acro{JO-PF}{joint opportunistic proportional fair}
\acro{OSI}{open systems interconnection}
\acro{PA}{power allocation}
\acro{PAIR}{\acs*{D2D} pair gain-based grouping}
\acro{PAPR}{peak-to-average power ratio}
\acro{P2P}{peer-to-peer}
\acro{PBCH}{physical broadcast channel}
\acro{PBS}{pico base station}        
\acro{PC}{power control}
\acro{PCI}{physical cell \acs*{ID}}
\acro{PDCP}{packet data convergence protocol}
\acro{PDF}{probability density function}
\acro{PDU}{protocol data unit}
\acro{PER}{packet error rate}
\acro{PF}{proportional fair}
\acro{PGRA}{probabilistic graph based resource allocation}
\acro{P-GW}{packet data network gateway}
\acro{PHY}{physical}
\acro{PIN}{positive-intrinsic-negative}
\acro{PL}{pathloss}
\acro{PLR}{packet loss ratio}
\acro{PPP}{Poisson point process}
\acro{PRABE}{power and resource allocation based on quality of experience}
\acro{PRB}{physical resource block}
\acro{PROJ}{projection-based grouping}
\acro{PSD}{power spectral density}
\acro{PSDe}{positive semi-definite}
\acro{ProSe}{proximity services}
\acro{PS}{packet scheduling}
\acro{PSO}{particle swarm optimization}
\acro{PSS}{primary synchronization signal}
\acro{PTAS}{polynomial-time approximation scheme}
\acro{PtP}{point-to-point}
\acro{PtmP}{point-to-multi-point}
\acro{PZF}{projected zero-forcing}
\acro{QAM}{quadrature amplitude modulation}
\acro{QHMLWDF}{queue-HOL-MLWDF}
\acro{QoE}{quality of experience}
\acro{QoS}{quality of service}
\acro{QPSK}{quadri-phase shift keying}
\acro{QSM}{queue-based satisfaction maximization}
\acro{QuaDRiGa}{quasi deterministic radio channel generator}
\acro{RACH}{random access}
\acro{RAISES}{reallocation-based assignment for improved spectral efficiency and satisfaction}
\acro{RAN}{radio access network}
\acro{RA}{resource allocation}
\acro{RAP}{rb assignment problem}
\acro{RATE}{rate-based}
\acro{RAU}{remote antenna unit}
\acro{RB}{resource block}
\acro{RBG}{resource block group}
\acro{REF}{reference grouping}
\acro{RET}{remote electrical tilt}
\acro{RF}{radio frequency}
\acro{RL}{reinforcement learning}
\acro{RLC}{radio link control}
\acro{RLF}{radio link failure}
\acro{RM}{rate maximization}
\acro{RMa}{rural macro}
\acro{RMJ-SNR}{region of the minimum joint snr}
\acro{RMEC}{rate maximization under experience constraints}
\acro{RMSE}{root-mean-square error}
\acro{RNC}{radio network controller}
\acro{RND}{random grouping}
\acro{RNN}{recurrent neural network}
\acro{RRA}{radio resource allocation}
\acro{RRM}{radio resource management}
\acro{RSCP}{received signal code power}
\acro{RSRP}{reference signal received power}
\acro{RSRQ}{reference signal received quality}
\acro{RR}{round robin}
\acro{RIS}{reconfigurable intelligent surface}
\acro{RRC}{radio resource control}
\acro{RSSI}{received signal strength indicator}
\acro{RT}{real time}
\acro{RTS}{request to send}
\acro{RU}{resource unit}
\acro{RUNE}{rudimentary network emulator}
\acro{RV}{random variable}
\acro{Rx}{receiver}
\acro{RZF}{regularized zero-forcing}
\acro{SA}{subcarrier assignment}
\acro{SAC}{session admission control}
\acro{sBS}{serving base station}
\acro{SC}{small cell}
\acro{SCI}{side control information}
\acro{SCG}{secondary cell group}
\acro{SCM}{spatial channel model}
\acro{SCS}{subcarrier spacing}
\acro{SC-FDMA}{single carrier - frequency division multiple access}
\acro{SCRV}{spatially consistent random variable}
\acro{SD}{soft dropping}
\acro{SDAP}{service	data adaptation protocol}
\acro{S-D}{source-destination}
\acro{SDPC}{soft dropping power control}
\acro{SDM}{space-division multiplexing}
\acro{SDMA}{space-division multiple access}
\acro{SE}{spectral efficience}
\acro{SF}{shadow fading}
\acro{SMDP}{semi-markov	decision problem}
\acro{SeNB}{secondary \acs*{ENB}}
\acro{SER}{symbol error rate}
\acro{SES}{simple exponential smoothing}
\acro{S-GW}{serving gateway}
\acro{SIC}{successive interference cancellation}
\acro{SINR}{signal to interference-plus-noise ratio}
\acro{SLNR}{signal to leakage-plus-noise ratio}
\acro{SNN}{strictly non-negative}
\acro{SI}{satisfaction indicator}
\acro{SIP}{session initiation protocol}
\acro{SISO}{single input single output}
\acro{SIMO}{single input multiple output}
\acro{SIR}{signal to interference ratio}
\acro{SM}{subcarrier matching}
\acro{SMA}{simple moving average}
\acro{SMSE}{spatial-mean-square error}
\acro{SN}{secondary node}
\acro{SNR}{signal to noise ratio}
\acro{SON}{self organizing networks}
\acro{SoA}{state-of-the-art}
\acro{SoS}{sum-of-sinusoids}
\acro{SORA}{satisfaction oriented resource allocation}
\acro{SORA-NRT}{satisfaction-oriented resource allocation for non-real time services}
\acro{SORA-RT}{satisfaction-oriented resource allocation for real time services}
\acro{SP}{signal processing}
\acro{SPF}{single-carrier proportional fair}
\acro{SR}{smart repeater}
\acro{ASR}{advanced \ac{SR}}
\acro{SRA}{sequential removal algorithm}
\acro{SRB1}{signaling radio bearer~1}
\acro{SRS}{sounding reference signal}
\acro{SRI}{\ac{SRS} resource indicator}
\acro{SSB}{synchronization signal block}
\acro{SSP}{small scale parameter}
\acro{SSS}{secondary synchronization signal}
\acro{ST}{spanning tree}
\acro{STAR}{simultaneously transmitting and reflecting}
\acro{SU-MIMO}{single-user multiple input multiple output}
\acro{SU}{single-user}
\acro{TA}{topology adaptation}
\acro{tBS}{target base station}
\acro{SVD}{singular value decomposition}
\acro{TCP}{transmission control protocol}
\acro{TDD}{time division duplex}
\acro{TDM}{time division multiplexing}
\acro{TDMA}{time division multiple access}
\acro{TDMed}{time division multiplexed}
\acro{TETRA}{terrestrial trunked radio}
\acro{TP}{transmit power}
\acro{TPC}{total power constraint}
\acro{TR}{technical report}
\acro{TS}{technical specification}
\acro{TTI}{transmission time interval}
\acro{TTR}{time-to-rendezvous}
\acro{TTT}{time-to-trigger}
\acro{TSM}{throughput-based satisfaction maximization}
\acro{TU}{typical urban}
\acro{TV}{television}
\acro{TCI}{transmission configuration indicator}
\acro{TVWS}{\acs*{TV} white space}
\acro{Tx}{transmitter}
\acro{UAV}{unmanned aerial vehicle}
\acro{UABSC}{user-assisted bearer split control}
\acro{UDP}{user datagram protocol}
\acro{UDN}{ultra-dense networks}
\acro{UE}{user equipment}
\acro{UBA}{\ac{UE}-\ac{BS} association}
\acro{UEPS}{urgency and efficiency-based packet scheduling}
\acro{UFC}{federal university of cear\'{a}}
\acro{ULA}{uniform linear array}
\acro{UL}{uplink}
\acro{UMa}{urban macro}
\acro{UMi}{urban micro}
\acro{UMTS}{universal mobile telecommunications system}
\acro{UP}{user plane}
\acro{UPA}{uniform planar array}
\acro{UPN}{user provided network}
\acro{URA}{uniform rectangular array}
\acro{URI}{uniform resource identifier}
\acro{URLLC}{ultra-reliable low-latency communications}
\acro{URM}{unconstrained rate maximization}
\acro{VBR}{variable bit rate}
\acro{VET}{variable electrical tilt}
\acro{VMR}{vehicle-mounted relay}
\acro{VR}{virtual resource}
\acro{VoIP}{voice over \acs*{IP}}
\acro{WCDMA}{wideband code division multiple access}
\acro{WF}{water-filling}
\acro{WID}{wireless infrastructure drone}
\acro{Wi-Fi}{wireless fidelity}
\acro{WiMAX}{worldwide interoperability for microwave access}
\acro{WINNER}{wireless world initiative new radio}
\acro{WLAN}{wireless local area network}
\acro{WMPF}{weighted multicarrier proportional fair}
\acro{WP}{work package}
\acro{WPF}{weighted proportional fair}
\acro{WSN}{wireless sensor network}
\acro{WWW}{world wide web}
\acro{WFQ}{weighted fair queuing}
\acro{XIXO}{(single or multiple) input (single or multiple) output}
\acro{XPR}{cross-polarization power ratio}
\acro{ZF}{zero-forcing}
\acro{ZMCSCG}{zero mean circularly symmetric complex gaussian}

\usepackage{siunitx}
\usepackage{datetime}
\usepackage{url}
\usepackage{tablefootnote}
\usepackage[caption=false,font=footnotesize]{subfig}

\usepackage[linesnumbered,ruled,vlined]{algorithm2e}

\usepackage{epstopdf}

\usepackage[hidelinks=true]{hyperref}

\usepackage{enumerate}

\usepackage[final]{changes} 
\definechangesauthor[name={Victor F. Monteiro}]{vfm}
\definechangesauthor[name={Tarcisio F. Maciel}]{tfm}
\setaddedmarkup{{\color{blue}#1}}
\setdeletedmarkup{{\color{red}\sout{#1}}}

\usepackage[referable,flushleft]{threeparttablex}
\AtBeginEnvironment{tablenotes}{\footnotesize} 

\DeclareMathAlphabet{\mathppl}{T1}{ppl}{m}{it}
\DeclareMathAlphabet{\mathphv}{T1}{phv}{m}{it}
\DeclareMathAlphabet{\mathpzc}{T1}{pzc}{m}{it}

\newcommand{\SecRef}[2][]{section#1~\ref{#2}}
\newcommand{\SubSecRef}[2][]{subsection#1~\ref{#2}}
\newcommand{\FigRef}[2][]{Fig.#1~\ref{#2}}

\newcommand{\AlgRef}[2][]{Algorithm#1~\ref{#2}}

\usepackage{standalone}
\usepackage{shapepar}

\usepackage{tikz}
\usetikzlibrary{arrows}
\usetikzlibrary{positioning}
\usetikzlibrary{shapes.misc}
\usetikzlibrary{shapes.geometric}
\usetikzlibrary{shapes.symbols}
\usetikzlibrary{external}

\usetikzlibrary{shadows}
\usetikzlibrary{backgrounds}
\usetikzlibrary{shapes}
\usetikzlibrary{shapes.multipart}
\usetikzlibrary{matrix}
\usetikzlibrary{intersections}
\usetikzlibrary{fit}
\usetikzlibrary{calc}
\usetikzlibrary{chains}
\usetikzlibrary{scopes}
\usetikzlibrary{decorations.pathreplacing}
\usetikzlibrary{decorations.text}
\usetikzlibrary{arrows.meta}
\usetikzlibrary{mindmap}
\usetikzlibrary{trees}
\usetikzlibrary{patterns}

\usepackage{pgfplots}
\usepgfplotslibrary{groupplots}

\pgfplotsset{%
	width=0.95\columnwidth,
	height=0.25\textheight, 
	compat=1.14,
	compat/show suggested version=false,
	filter discard warning=false,
	tick label style={font=\footnotesize},
	label style={font=\footnotesize},
	every axis label={font=\footnotesize},
	grid=major,
	grid style={dashed,gray!30},
	cycle list shift=0,
	enlargelimits=false,
	legend style={%
		font=\footnotesize,
		legend cell align=left,
		nodes={inner xsep=2pt,inner ysep=1pt,text depth=0.15em},
	},
}
\usepackage{pgfplotstable}

\newtoggle{tikzexternalize}
\togglefalse{tikzexternalize}

\iftoggle{tikzexternalize}{
	\immediate\write18{mkdir -p ./tikz-external-figs/}
	\usetikzlibrary{external}
	\tikzexternalize[prefix=./tikz-external-figs/,mode=list and make]
	\tikzset{external/system call={pdflatex \tikzexternalcheckshellescape -halt-on-error -interaction=batchmode -jobname "\image" "\texsource"}}
}{
}

\AtBeginDocument{%
	\abovedisplayskip 1.5ex plus 4pt minus 2pt
	\belowdisplayskip \abovedisplayskip
	\abovedisplayshortskip 0pt plus 4pt
	\belowdisplayshortskip 1.5ex plus 4pt minus 2pt
}

\pgfplotsset{common line style/.style={line width=1pt}}

\pgfplotsset{every axis plot post/.append style={
    every mark/.append style={scale=1.5}
}}

\pgfplotsset{common plots axis options/.style={
	every axis/.append style={
	  legend style={fill=gray!5, fill opacity=0.85, text opacity=1}
	},
	width=\columnwidth,
	height=0.4\columnwidth,
	grid=both,
	filter discard warning=false,
	tick label style={font=\footnotesize},
	label style={font=\footnotesize},
	every axis label={font=\footnotesize},
	grid=major,
	grid style={dashed,gray!30},
	cycle list shift=0,
	enlargelimits={true,abs value=1pt},
	ylabel shift = -0.5ex,
	legend style={%
		font=\scriptsize,
		legend cell align=left,
		nodes={inner xsep=2pt,inner ysep=1pt,text depth=0.15em},
	},
	}
}

\pgfplotsset{bar axis options/.style={
		common plots axis options,
		ybar=1pt,
		bar width = 3pt,
		enlarge x limits={true,abs value=5pt},
}}

\pgfplotsset{barLoad axis options/.style={
		common plots axis options,
		ybar=7pt,
		bar width=35pt,
		enlarge x limits={true,abs value=35pt},
}}

\pgfplotsset{mcs axis options/.style={
		bar axis options,
		width=\columnwidth,
		height=0.4\columnwidth,
		ymin = 0, ymax = 100,
		xtick={0, 1, ..., 15},
		xticklabels={Total, MCS 1, MCS 2, MCS 3, MCS 4, MCS 5, MCS 6, MCS 7, MCS 8, MCS 9, MCS 10, MCS 11, MCS 12, MCS 13, MCS 14, MCS 15},
		x tick label style={
			font=\tiny,
			xshift = 1ex,
			rotate=45,
			anchor=east,
		},
}}

\pgfplotsset{timeConnec axis options/.style={
		barLoad axis options,
		width=\columnwidth,
		height=0.4\columnwidth,
		ymin = 0, ymax = 100,
		xtick={1, 2, 3},
		xticklabels={Donor 1, Donor 2, Donor 3},
		x tick label style={
			font=\small,
			xshift=5ex,
			yshift=-2ex,
			rotate=0,
			anchor=east,
		},
}}

\pgfplotsset{common marker style/.style={
		mark repeat = 10,
		mark size = 1pt,
		mark options={solid},
}}

\pgfplotsset{scenario311Dir style/.style={
	common line style,
	common marker style,
	MediumSeaGreen,
	mark=x,
	mark size=1.3pt,
}}

\pgfplotsset{scenario322Dir style/.style={
	common line style,
	common marker style,
	Red,
	mark=*,
	mark size=0.8pt,
}}

\pgfplotsset{scenario01Dir style/.style={
    common line style,
    common marker style,
    DodgerBlue,
    mark=triangle*,
    mark size=1pt,
}}

\pgfplotsset{scenario02Dir style/.style={
		common line style,
		common marker style,
		Orange,
		mark=square*,
		mark size=0.7pt,
}}

\pgfplotsset{scenario311Ncr style/.style={
		dashed,
		common marker style,
		MediumSeaGreen,
		mark=x,
		mark size=1.4pt,
}}

\pgfplotsset{scenario322Ncr style/.style={
		dashed,
		common marker style,
		Red,
		mark=*,
		mark size=1pt,
}}

\pgfplotsset{T1 style/.style={
		common line style,
		common marker style,
		DodgerBlue,
		mark=triangle*,
}}

\pgfplotsset{T1LB style/.style={
		dashed,
		common marker style,
		DodgerBlue,
		mark=triangle*,
		line width=1pt,
}}

\pgfplotsset{T2 style/.style={
		common line style,
		common marker style,
		Orange,
		mark=square*,
}}

\pgfplotsset{T2LB style/.style={
		dashed,
		common marker style,
		Orange,
		mark=square*,
		line width=1pt,
}}

\pgfplotsset{T3 style/.style={
		common line style,
		common marker style,
		MediumSeaGreen,
		mark=*,
}}

\pgfplotsset{T3LB style/.style={
		dashed,
		common marker style,
		MediumSeaGreen,
		mark=*,
		line width=1pt
}}

\pgfplotsset{standard style/.style={common line style, solid}}
\pgfplotsset{load style/.style={common line style, dashed}}

\pgfplotsset{ack bar style/.style={
		DodgerBlue, fill
}}

\pgfplotsset{nack bar style/.style={
		Red, fill
}}

\pgfplotsset{node1 barLoad style/.style={
		DodgerBlue, fill
}}

\pgfplotsset{node2 barLoad style/.style={
		Red, fill, pattern=north east lines,
		pattern color=red
}}

 \def\plotsDataPath{figs/plots/data}

\begin{document}
\title{Load Balancing-based Topology Adaptation for \\Integrated Access and Backhaul Networks
}

\author{
Raul Victor de O. Paiva, Fco. Italo G. Carvalho, Fco. Rafael M. Lima, Victor F. Monteiro, \\ Diego A. Sousa, Darlan C. Moreira,  Tarcisio F. Maciel and Behrooz Makki
\thanks{Behrooz Makki is with Ericsson Research, Sweden. The other authors are with the Wireless Telecommunications Research Group (GTEL), Federal University of Cear\'{a} (UFC), Fortaleza, Cear\'{a}, Brazil. Diego A. Sousa is also with Federal Institute of Education, Science, and Technology of Cear\'{a} (IFCE), Paracuru, Brazil. This work was supported by Ericsson Research, Sweden, and Ericsson Innovation Center, Brazil, under UFC.51 Technical Cooperation Contract Ericsson/UFC. The work of Victor F. Monteiro was supported by CNPq under Grant 308267/2022-2. The work of Tarcisio F. Maciel was supported by CNPq under Grant 312471/2021-1. The work of Francisco R. M. Lima was supported by FUNCAP (edital BPI) under Grant BP4-0172-00245.01.00/20.}%
}

\maketitle

\begin{abstract}
	%
	\Ac{IAB} technology is a \deleted{cost-effective and }flexible solution for network densification. %
	\ac{IAB} nodes can also be deployed in moving nodes such as buses and trains, i.e., \ac{mIAB}. %
	As \ac{mIAB} nodes can move around the coverage area, the connection between \ac{mIAB} nodes and their parent macro \acp{BS}, \ac{IAB} donor, is sometimes required to change in order to keep an acceptable backhaul link, the so called \ac{TA}. %
	The change from one \ac{IAB} donor to another may strongly impact \deleted{on} the system load distribution, possibly causing unsatisfactory backhaul service due to the lack of radio resources. %
	Based on this, \ac{TA} should consider both backhaul link quality and traffic load. %
	In this work\added{,} we propose\deleted{d} a load balancing algorithm based on \ac{TA} for \ac{IAB} networks, and compare\deleted{d} it with an approach in which \ac{TA} is triggered based on \ac{RSRP} only. %
	The results show that \replaced{our}{the} proposed algorithm
	improves the passengers worst connections throughput in \ac{UL} and, more modestly, also in \ac{DL}, without impairing the pedestrian \ac{QoS} significantly. %
\end{abstract}

\begin{IEEEkeywords}
	\ac{mIAB}, topology adaptation, load balancing.
\end{IEEEkeywords}

%
\IEEEpeerreviewmaketitle
\acresetall

\section{Introduction}
\label{SEC:intro}

The use of the large available bands in \ac{mmWave} frequencies in \ac{5G} networks allows the fulfillment of \ac{QoS} requirements for new data-hungry services. %
However, that part of the spectrum suffers from severe path and penetration losses~\cite{Flamini2022,Ayoubi2022}. %
To overcome these drawbacks, one \added{envisioned} solution \deleted{being envisioned} is the densification of the network with \ac{IAB} nodes. %

\added{In summary, an \ac{IAB} node can be seen as a regular \ac{gNB} in which the backhaul is wireless, being served by another network node, called \ac{IAB} donor}. 
\ac{IAB} nodes enable deploying new infrastructure avoiding the high costs and time consumption of traditional wired installations and overcoming possible limitations on trenching, such as in historical places~\cite{Polese2020,Madapatha2020,Monteiro2022_1,Teyeb2019}. %

In Release~18~\cite{3gpp.22.839}, it was investigated applications and technical requirements for \ac{5G} networks that use \acp{VMR} that provide service to
onboard passengers and surrounding pedestrians. %
Additionally, \acp{VMR} are designed to facilitate seamless connectivity despite the mobility of the \acp{UE} and the \ac{VMR} itself. %
For \ac{IAB} networks, \acp{VMR} can be identified as \ac{mIAB}~\cite{3gpp.22.839}. %

\Ac{TA}, in this context, refers to the migration of an \ac{IAB} node from an \ac{IAB} donor to another one, which is triggered when the link between an \ac{IAB} node and its current donors weakens, due to mobility and/or obstacles~\cite{3gpp.38.874}. %
As \ac{mIAB} nodes \replaced{move,}{moves} \deleted{the mobile network coverage area carrying passenger \acp{UE},} the traffic load in the system becomes more dynamic and heterogeneous. %
Specifically, \replaced{in a given instant of time, one}{while a} macro \ac{BS} may serve only a few pedestrians, \added{while} other macro \acp{BS} may serve several \ac{mIAB} nodes \replaced{through their}{in} backhaul as well as \added{direct} pedestrians. %
Thus, depending on the system load distribution, not all \ac{IAB} donors can provide satisfactory backhaul service to a coming \ac{mIAB} node, due to the lack of available radio resources. %

The authors in \cite{Monteiro2022} surveyed several works on \ac{IAB} and \ac{mIAB} and presented a detailed performance evaluation in order to assess the gains of \ac{mIAB} over conventional deployments. %
It was shown that the load variability among \ac{IAB} donors is in general high with \ac{mIAB} nodes. %
The authors then discussed the need of \ac{TA} strategies that consider \ac{IAB} donor load to avoid overloading \ac{IAB} donors. %

In \cite{Choi2021}, the authors proposed a method to distribute the load among fixed \ac{IAB} nodes efficiently by sharing their load information and data processing capabilities. %
The proposed method outperformed traditional approaches based on metrics such as \ac{RSRP} in terms of average throughput and number of served \acp{UE}. %
Motivated by the conclusions of~\cite{Monteiro2022, Choi2021} and use cases involving load balancing in Release 18~\cite{3gpp.22.839}, we investigate load balancing based \ac{TA} in this paper. %

\comment{VEJA QUE O TEMPO VERBAL DAS 3 REFERENCIAS CITADAS A SEGUIR ESTÁ DIFERENTE ENTRE ELAS: ``the authors proposed'', ``it is proposed'', ``which addresses'' ``The work focused''. OU COLOCA TD NO PRESENTE OU TD NO PASSADO}

\deleted{The rest of this letter is organized as follows.} %
\deleted{In sections \ref{SEC:IAB} and \ref{SEC:system_model} we present a brief review of the \ac{IAB} technology and the assumed system modeling, respectively.} %
\deleted{The simulation assumptions and the results are reported in~\SecRef{SEC:performance_evaluation}.} %
\deleted{The conclusions and perspectives are presented in~\SecRef{SEC:conclusions}.} %

\added{In this letter, we propose a \ac{TA} strategy that takes into account not only the channel strength but also the system load in a \ac{mIAB} network.} %
\added{Results show that our proposed solution outperforms the standard \ac{TA} solution, leading to improved throughput for passengers without significantly deteriorating the pedestrians' \ac{QoS}.} %

\section{Mobile Integrated Access and Backhaul}
\label{SEC:IAB}

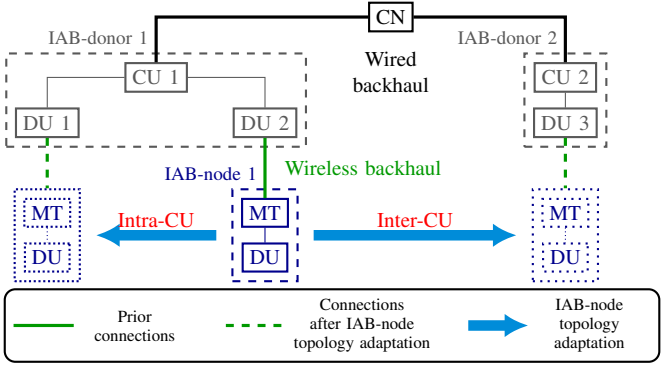
\begin{figure}[t]
	\centering
\begin{tikzpicture}[scale = 0.8, transform shape, every node/.append style={thick}]
  \colorlet{DonorColor}{gray!70!black}
  \colorlet{IABNodeColor}{blue!55!black}
  \colorlet{Wireless Backhaul Color}{green!60!black}
  \definecolor{topology adaptation color}{rgb}{00,0.55,.85}

  \tikzset{text box/.style={text width = 2cm, font=\footnotesize, align=center, inner sep = 0}}
  \tikzset{legend box/.style={text width = 2cm, font=\footnotesize, align=center, text=black, inner sep = 0}}

  \tikzset{node box/.style={draw, minimum width=0.50em, minimum height=0.99em}}
  \tikzset{backhaul/.style={very thick}}
  \tikzset{topology adaptation style/.style={dashed}}
  \tikzset{topology arrow line/.style={topology adaptation color, line width=1.4mm,arrows={-Latex[width=9pt, length=9pt]}}}


  \node[node box] (CN) at (0, 0) {CN};

  { [DonorColor]
    \node[node box, below left = 0.5cm and 3cm of CN] (CU 1) {CU 1};
    \node[node box, below left = 0.25cm and 0.75cm of CU 1] (DU 1) {DU 1};
    \node[node box, below right = 0.25cm and 0.75cm of CU 1] (DU 2) {DU 2};
    \node[draw, dashed, fit=(CU 1)(DU 1)(DU 2), label={[above left, font=\small]IAB-donor 1}] (Donor1) {};

    \node[node box, below right = 0.5cm and 2cm of CN](CU 2) {CU 2};
    \node[node box, below = 0.25cm of CU 2] (DU 3) {DU 3};
    \node[draw, dashed, fit=(CU 2)(DU 3), label={[above left, font=\small]IAB-donor 2}] (Donor2) {};

    \draw (CU 1) -| (DU 1);
    \draw (CU 1) -| (DU 2);
    \draw (CU 2) -- (DU 3);
  }

  { [backhaul]
      \draw (CN) -| (CU 1);
      \draw (CN) -| (CU 2);
      \node[below=2mm of CN, text width=4em, text centered] {Wired backhaul};
  }

  { [IABNodeColor]
      \node[node box, below = 1cm of DU 2] (Node1-MT) {MT};
      \node[node box, below = 0.25cm of Node1-MT] (Node1-DU) {DU};
      \node[draw, dashed, fit=(Node1-MT)(Node1-DU), label={[above left, font=\small]IAB-node 1}] (Node1) {};

      \draw (Node1-MT) -- (Node1-DU);

      { [every node/.append style={node box, topology adaptation style}]

      \node[node box, densely dotted, below = 1cm of DU 1] (left Node1-MT) {MT};
      \node[node box, densely dotted, below = 0.25cm of left Node1-MT] (left Node1-DU) {DU};
      \node[draw, densely dotted, fit=(left Node1-MT)(left Node1-DU)] (moved left Node1) {};
      \draw[densely dotted] (left Node1-MT) -- (left Node1-DU);

      \node[node box, dotted, below = 1cm of DU 3] (right Node1-MT) {MT};
      \node[node box, dotted, below = 0.25cm of right Node1-MT] (right Node1-DU) {DU};
      \node[draw, dotted, fit=(right Node1-MT)(right Node1-DU)] (moved right Node1) {};
      \draw[densely dotted] (right Node1-MT) -- (right Node1-DU);


      }
  }

  { [backhaul, Wireless Backhaul Color]
    \draw (DU 2) -- node[right, text width=3cm, text centered] {Wireless backhaul} (Node1-MT);

    { [topology adaptation style]
      \draw (DU 1) -- (moved left Node1);
      \draw (DU 3) -- (moved right Node1);
    }
  }

  { [topology arrow line, shorten <=2mm, shorten >=2mm]
    \draw (Node1) -- node[above, red] {Intra-CU} (moved left Node1.east);
    \draw (Node1) -- node[above, red] {Inter-CU} (moved right Node1.west);
  }

  \def\legendlinewidth{1.0cm}
  \coordinate[below = 1.5cm of moved left Node1.west] (legend);
  \draw[backhaul, Wireless Backhaul Color] (legend) -- ++(\legendlinewidth, 0) node[right, legend box] (prior connections) {Prior\\connections};

  \draw[backhaul, Wireless Backhaul Color, topology adaptation style] (prior connections.east) ++(5mm, 0) -- ++(\legendlinewidth, 0) node [right, legend box, text width = 2.5cm] (conn topology adaptation) {Connections after IAB-node topology adaptation};

  \draw[topology arrow line] (conn topology adaptation.east) ++(5mm, 0) -- ++(\legendlinewidth, 0) node[right, legend box] (iab topology adaptation) {IAB-node\\topology adaptation};

  \node[rounded corners, draw, fit=(legend)(prior connections)(conn topology adaptation)(iab topology adaptation)] (legends) {};
\end{tikzpicture}
	\caption{Illustration of \ac{IAB} \ac{NR} architecture with \ac{IAB} donor and nodes main components and topology adaptation~\cite{Monteiro2022}.}
	\label{FIG:Architecture-IAB-topology-adaptation}
\end{figure}

The \ac{IAB} architecture, illustrated in \FigRef{FIG:Architecture-IAB-topology-adaptation}, was designed to support multiple wireless backhaul hops in \ac{NR}. %
Therein, the blue boxes represent the \ac{IAB} nodes. These nodes can connect to each other wireless\added{ly}. %
The \ac{IAB} donors, shown as gray boxes, have a wired backhaul and provide wireless backhaul to an \ac{IAB} node. %

The \ac{IAB} nodes can be \replaced{split}{separated} into \ac{DU} and \ac{MT}~\cite{3gpp.38.300c}. %
The \ac{MT} component is responsible for managing the radio signal used to connect to a parent node that can be an \ac{IAB} donor or an \ac{IAB} node. 
The \ac{DU} component serves as the \ac{NR} interface for both \acp{UE} and the \ac{MT} part of a child \ac{IAB} node. %
The \ac{IAB} donors, with wired backhaul to the \ac{CN}, have two parts: \ac{CU} and \ac{DU}. %
Lower protocol layers are responsibility of the \acp{DU}, while the upper protocol layers are provided by the \acp{CU}~\cite{3gpp.38.300c}. %
The aforementioned split permits time-critical functionalities to be conducted in the \ac{DU} near the served nodes, while others are handled with more processing power in the \ac{CU}~\cite{Madapatha2020}.

Wireless backhaul allows for the deployment of mobile cells, the \ac{mIAB} nodes, that can provide uninterrupted cellular services for moving \acp{UE}~\cite{Monteiro2022}. %
The main use cases for \ac{mIAB} technology consist in the service provision for both passengers onboard and pedestrians in the adjacency of the vehicle~\cite{3gpp.22.839}. %
For buses, the outer antennas on top correspond to the \ac{MT}, while the inner antennas correspond to the \ac{DU}. %


Concerning \ac{TA}, an \ac{IAB} node might need to switch to a different parent node after initial setup. %
This can happen, for example, if the connection to its current parent weakens due to movement or obstructions between them~\cite{3gpp.38.874}. %
In fixed \ac{IAB} scenarios, \ac{TA} due to poor channel quality is less frequent, as nodes are stationary\added{ and well planned}. %
In contrast, in \ac{mIAB} scenarios, \ac{TA} mainly occurs due to channel variations caused by the movement of transmitters and receivers~\cite{3gpp.38.874}. %

Figure~\ref{FIG:Architecture-IAB-topology-adaptation} shows a \ac{TA} illustration for \ac{IAB} node 1 in a scenario with two \ac{IAB} donors (donors 1 and 2)\deleted{ and three other fixed nodes (nodes 1, 2, and 3)}. %
%
%
Assuming \ac{IAB} node 1 as an \ac{mIAB}, we can see that it has \replaced{donor 1 (through \ac{DU} 2)}{node 2} as parent node and needs to switch to another parent node due to poor backhaul link quality. %
\replaced{Donor 1 (through \ac{DU} 1) and donor 2}{Nodes 1 and 3} are candidate parent nodes. %
In this case, \replaced{donor 1 through DU 1}{node 1} may provide a connection with the same \ac{IAB} donor-\ac{CU} (intra-\ac{CU}), while \replaced{donor 2}{node 3} may provide a different \ac{IAB} donor-\ac{CU} (inter-\ac{CU}). %
The inter-\ac{CU} case involves handover requests~\cite{3gpp.38.874}. %


\section{System Model and Load-Aware Topology Adaptation}
\label{SEC:system_model}
\begin{figure}[t]
	\centering
	\makeatletter%
	\if@twocolumn%
		\includegraphics[width=0.90\columnwidth]{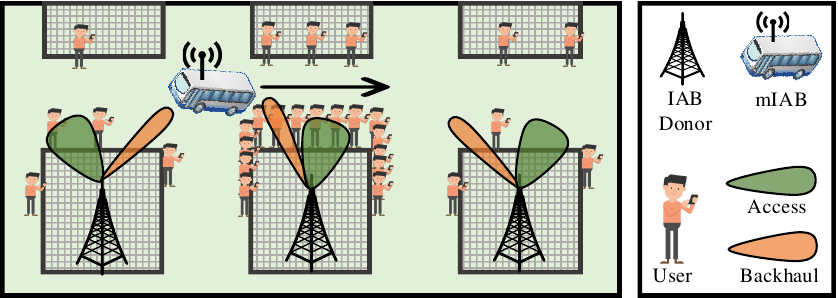}
	\else
		\includegraphics[width=0.45\columnwidth]{figs/scenario_madrid_reduced.pdf}
	\fi
	\makeatother
	\caption{Scenario of interest.}
	\label{FIG:scenario}
\end{figure}

%
In~\FigRef{FIG:scenario}, the proposed study scenario is shown. %
It considers a single bus equipped with an \ac{mIAB}, which is connected to an \ac{IAB} donor and, as the time passes, gets closer to two other donors, the closest (center) one being overloaded. %

Conventionally, the \ac{mIAB} node performs \ac{TA} to the \ac{IAB} donor that provides the strongest backhaul link quality, e.g., highest \ac{RSRP}, which is usually the closest one. %
However, in the considered scenario, the choice of the \ac{IAB} donor with stronger \ac{RSRP} may degrade the connection of of onboard \acp{UE}, since the the closest \ac{IAB} donor is overloaded. %
Thus, few radio resources would be left for backhaul connection. %
%

\FloatBarrier

A solution for this situation is to avoid the connection of the incoming
\ac{mIAB} node, i.e., \ac{TA}, to the overloaded \ac{IAB} donor and connecting it to another, less
loaded, \ac{IAB} donor instead. %
Therefore, to that end, the following~\AlgRef{ALG:algorithm_load_balancing} is proposed. %
It aims at selecting a candidate \ac{IAB} donor to which to handover the \ac{mIAB} node. %
It takes into account not only the \ac{RSRP} of a candidate \ac{IAB} donor, but also its load, represented by the quantity of buffered bits at the \ac{IAB} donors. %
Specifically, besides of having an \ac{RSRP} value greater than a selected minimum value (line~\ref{alg:RSRP_threshold_test}) and an \ac{RSRP} not significantly smaller than the \ac{RSRP} value of the current \ac{IAB} donor (line~\ref{RSRP_eval}), the candidate serving \ac{IAB} donor must have a traffic load smaller than the current serving one (line~\ref{alg:load_test}). %

\begin{algorithm}[!bp]\footnotesize
	\SetAlgoNlRelativeSize{0}
	\KwIn{List of~\ac{IAB} donors: {\fontfamily{pcr}\selectfont \textit{donors}}}
	\KwOut{Updated parent~\ac{IAB} donor: {\fontfamily{pcr}\selectfont \textit{parent}}}
	\ForEach{{\fontfamily{pcr}\selectfont candidate} in {\fontfamily{pcr}\selectfont donors}}{
		\If{{\fontfamily{pcr}\selectfont candidate} is not {\fontfamily{pcr}\selectfont parent}}{
			\If{{\fontfamily{pcr}\selectfont candidate.bits} $<$ {\fontfamily{pcr}\selectfont parent.bits}}{ \label{alg:load_test}
				\If{{\fontfamily{pcr}\selectfont candidate.RSRP} $>$ {\fontfamily{pcr}\selectfont minRSRP}}{ \label{alg:RSRP_threshold_test}
					\If{{\fontfamily{pcr}\selectfont candidate.RSRP} $>$ {\fontfamily{pcr}\selectfont parent.RSRP} - {\fontfamily{pcr}\selectfont minRSRPDiff}}{ \label{RSRP_eval}
						{\fontfamily{pcr}\selectfont \textit{parent}} $\leftarrow$ {\fontfamily{pcr}\selectfont \textit{candidate}}\;
					}
				}
			}
		}
	}
	\caption{Update Parent Base Station}
	\label{ALG:algorithm_load_balancing}
\end{algorithm}
%

The main objective of this work is to compare a standard \ac{TA} algorithm~\cite{Monteiro2022} which considers only the \ac{RSRP} value as activation criterion, hereafter called standard algorithm, against the proposed~\AlgRef{ALG:algorithm_load_balancing}. %

\section{Performance Evaluation}
\label{SEC:performance_evaluation}
%
This section presents a performance comparison between the standard and the proposed \ac{TA} algorithms. %
The main simulation parameters are presented in~\SubSecRef{SUBSEC:simulation_assumptions} and the results are discussed in~\SubSecRef{SUBSEC:simulation_results}. %

\subsection{Simulation Assumptions}
\label{SUBSEC:simulation_assumptions}

The considered scenario has one bus that serves 6 passengers through its \ac{IAB} node, which is traveling with a constant speed of \SI{50}{km/h} on a trajectory that passes in front of 3 building blocks modeled using a simplified Madrid Grid~\cite{METIS:D6.1:2013,Sui2015} (see~\FigRef{FIG:scenario}). %
%
%
One \ac{IAB} donor is placed at the end of the opposing side of each block, from where the \ac{IAB} node passes. %
40 pedestrians moving with a speed of \SI{3}{km/h} are connected to the central \ac{IAB} donor, while the other \ac{IAB} donors serve only 5 pedestrians, each. %
We consider 22 \acp{RB}, that are shared by all the network nodes. %
Also, a \ac{CBR} traffic model is assumed, generating packets with size \SI{4,096}{bits} each \SI{1}{ms}. %
The values of the simulation parameters are presented in Tables~\ref{TABLE:Entities-characteristics} and~\ref{TABLE:Simul_Param}. %

\begin{table}[t]
    \scriptsize
    \centering
    \caption{Entities characteristics.}
    \label{TABLE:Entities-characteristics}
    \begin{tabular}{llllll}
        \toprule
        \multirow{2}{*}{\bf Parameter} & \multirow{2}{*}{\bf \ac{IAB} donor} & \multicolumn{2}{l}{\bf\hspace*{1em} \ac{mIAB} node} & \multicolumn{2}{l}{\bf\hspace*{2.2em} \ac{UE}} \\ \cmidrule(lr){3-4} \cmidrule(l){5-6}
        & & \textbf{\acs{DU}} & \textbf{\acs{MT}} & \textbf{Ped.} & \textbf{Pass.} \\
        \midrule
        Height & \SI{25}{\meter} & \SI{2.5}{\meter} & \SI{3.5}{\meter} & \SI{1.5}{\meter} & \SI{1.8}{\meter} \\ 
        Transmit power & \SI{35}{\decibel m} & \SI{24}{\decibel m} & \SI{24}{\decibel m} & \SI{24}{\decibel m} & \SI{24}{\decibel m} \\ 
        Antenna tilt & $12^{\circ}$ & $4^{\circ}$ & $0^{\circ}$ & $0^{\circ}$ & $0^{\circ}$ \\ 
        Antenna array & ULA $64$ & URA $8\!\times\! 8$ & ULA $64$ & \begin{tabular}[c]{@{}l@{}} Single \\ antenna \end{tabular} & \begin{tabular}[c]{@{}l@{}} Single \\ antenna \end{tabular} \\ 
        \begin{tabular}[c]{@{}l@{}}Antenna element \\ pattern \end{tabular} & \acs{3GPP}~\cite{3gpp.38.901} & \acs{3GPP}~\cite{3gpp.38.901} &  Omni & Omni & Omni \\ 
        \begin{tabular}[c]{@{}l@{}}Max. antenna \\ element gain \end{tabular} & \SI{8}{\decibel i} & \SI{8}{\decibel i} & \SI{0}{\decibel i} & \SI{0}{\decibel i} & \SI{0}{\decibel i} \\ 
        Speed & \SI{0}{km/h} & \SI{50}{km/h} & \SI{50}{km/h} & \SI{3}{km/h} & \SI{50}{km/h} \\
        \bottomrule
    \end{tabular}
\end{table}

\begin{table}[t]
	\centering
	\scriptsize
	\caption{Simulation parameters.}
	\label{TABLE:Simul_Param}
	\begin{tabularx}{\columnwidth}{lX}
		\toprule
		\textbf{Parameter}                          & \textbf{Value}                                                                  \\
		\midrule
		Layout                                      & Simplified Madrid grid~\cite{METIS:D6.1:2013,Sui2015}                           \\
		Carrier frequency                           & \SI{28}{\GHz}                                                                   \\
		Subcarrier spacing                          & \SI{60}{\kHz}                                                                   \\
		Number of subcarriers per \acs{RB}          & $12$                                                                            \\
		Number of \acsp{RB}                         & $22$                                                                            \\
		Slot duration                               & \SI{0.25}{\ms}                                                                  \\
		\acs{OFDM} symbols per slot                 & $14$                                                                            \\
		Channel generation procedure                & As described in~\cite[Fig.~7.6.4-1]{3gpp.38.901}                                \\
		Path loss                                   & As described in~\cite[Table 7.4.1-1]{3gpp.38.901}                               \\
		Fast fading                                 & As described in~\cite[Sec.7.5]{3gpp.38.901} and \cite[Table~7.5-6]{3gpp.38.901} \\
		\acs{AWGN} power per subcarrier             & \SI{-174}{dBm}                                                                  \\
		Noise figure                                & \SI{9}{\decibel}                                                                \\
		Mobility model                              & Pedestrian and Vehicular~\cite{3gpp.37.885}                                     \\
		Number of buses                             & 1                                                                               \\
		Passengers + pedestrians                    & 56                                                                              \\
		\acs{CBR} traffic packet size               & $4,096$ bits                                                                    \\
		\acs{CBR} traffic packet inter-arrival time & \SI{4}{slots}                                                                   \\
		\bottomrule
	\end{tabularx}
\end{table}

In order to avoid self-interference, \ac{IAB}-\ac{DU} and \ac{IAB}-\ac{MT} operate in \ac{HD}, meaning that they cannot transmit while the other is receiving, i.e., they operate together in the same direction, either both transmitting or receiving. %
%


\subsection{Simulation Results}
\label{SUBSEC:simulation_results}

This section presents: 1) the duration of the \ac{mIAB} node’s connection to each \ac{IAB} donor; 2) the time evolution of buffer size of access and backhaul links; and 3) the \ac{UL} and \ac{DL} throughput for pedestrians and passengers. 

\subsubsection{\ac{mIAB} node connection \replaced{time}{times}}
\label{SUBSUBSEC:mIAB_connection_times}

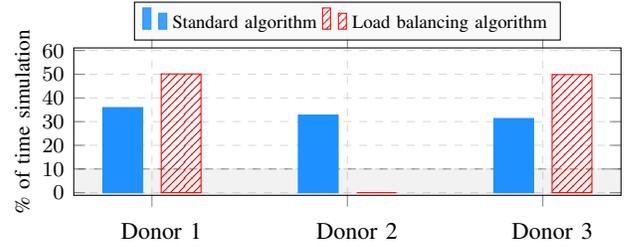
\begin{figure}[t]
	\centering
		\begin{tikzpicture}
		\begin{axis}[
            timeConnec axis options,
			ylabel=\% of time simulation,
			ymax = 60,
			ytick = {0,10,...,60},
            bar width = 15px,
            enlarge x limits=0.2,
			legend style={
		    	at = {(0.5, 1.05)},
                anchor = south,
		    	legend columns = 2
			}
		]

			\pgfplotstableread [col sep=comma]
			{\plotsDataPath/time_connection_data_LB_false.csv}\tableDataFalse

            \pgfplotstableread [col sep=comma]
            {\plotsDataPath/time_connection_data_LB_true.csv}\tableDataTrue

			\addlegendimage{node1 barLoad style}
			\addlegendentry{Standard algorithm}
            \addlegendimage{node2 barLoad style}
            \addlegendentry{Load balancing algorithm}

			\draw[black, opacity=0.4, fill=blue,sharp plot,dashed] (-1, 10) -- (16, 10) ;
			\fill[gray, opacity=0.1] (-1, -1) -- (-1, 10) -- (16, 10) -- (16, -1) -- cycle;

			\addplot[node1 barLoad style]
			table[x=donor, y=time] from \tableDataFalse;

            \addplot[node2 barLoad style]
            table[x=donor, y=time] from \tableDataTrue;

		\end{axis}
	\end{tikzpicture}
	\caption{Total time the \ac{mIAB} remains connected with donors 1, 2 and 3 with the standard and load balancing algorithms.}
	\label{FIG:time_connection}
\end{figure}

In \FigRef{FIG:time_connection}, we present the total time the \ac{mIAB} node (bus) is connected with each \ac{IAB} donor, with and without \deleted{the }usage of the proposed the load balancing \ac{TA}. %
We can see that using the standard \ac{TA} algorithm~\cite{Monteiro2022}, the \ac{mIAB} node remains connected to each \ac{IAB} donor approximatelly 33\% of the simulation time, as expected\deleted{,} considering the \ac{IAB} node constant speed and the grid symmetry. %
When~\AlgRef{ALG:algorithm_load_balancing} is used, there is no connection between the \ac{mIAB} node and the overloaded central block \ac{IAB} donor. %
This means the algorithm effectively accounts for the overload in the \ac{TA} process, improving \ac{QoS} for passengers by ensuring more resources are available in its serving \ac{IAB} donor. %

%

\subsubsection{Transmission Buffer Load}
\label{SUBSUBSEC:bits_load_in_transmission_buffer}

\begin{figure}[tbp]
	\centering
	   	\begin{tikzpicture}
        \begin{groupplot}[
           group style={
               group size = 1 by 2,
               xlabels at=edge bottom,
               xticklabels at=edge bottom,
               vertical sep = 5pt,
           },
           xlabel=\% of simulation time,
           xtick = {0,10.0,...,100.0},
           xmin=-0.001, xmax = 100.0,
        ]

        \pgfplotstableread [col sep=comma] {\plotsDataPath/buffer_UL_mt2donor_KBits_decimated.csv}\tableDataMTDonor
        \pgfplotstableread [col sep=comma] {\plotsDataPath/buffer_DL_donor2mt_KBits_decimated.csv}\tableDataDonorMT
        \pgfplotstableread [col sep=comma] {\plotsDataPath/buffer_DL_node2ue_KBits_decimated.csv}\tableDataNodeUE

        \nextgroupplot[
           common plots axis options,
           ymin=0, ymax = 25,
           ytick = {0,5.0,...,25.0},
           legend style={
               at = {(0.5, 1.05)},
               anchor = south,
               legend columns = 3,
           }
        ]

            \addlegendimage{scenario01Dir style, mark size = 1.5pt, only marks}
            \addlegendentry{Standard algorithm}

            \addlegendimage{scenario322Dir style, mark size = 1.5pt, only marks}
            \addlegendentry{Load balancing algorithm}

            \addlegendimage{white}
            \addlegendentry{}

            \addlegendimage{common line style, solid}
            \addlegendentry{Donor}

            \addlegendimage{common line style, dashed}
            \addlegendentry{\acs{mIAB} DU}

            \addlegendimage{common line style, dotted}
            \addlegendentry{\acs{mIAB} MT}

            \addplot[scenario01Dir style, dotted]
            table[x=time_percentage, y expr = \thisrow{LB_False} * 1e-3] from \tableDataMTDonor;

            \addplot[scenario322Dir style, dotted]
            table[x=time_percentage, y expr = \thisrow{LB_True} * 1e-3] from \tableDataMTDonor;

        \nextgroupplot[
           common plots axis options,
           ymin=0, ymax = 4,
           ylabel=No bits Tx buffer (MBits),
           ylabel shift = 1,
           every axis y label/.append style={at=(ticklabel cs:1.1)},
        ]

            \addplot[scenario01Dir style]
            table[x=time_percentage, y expr = \thisrow{LB_False} * 1e-3] from \tableDataDonorMT;

            \addplot[scenario322Dir style]
            table[x=time_percentage, y expr = \thisrow{LB_True} * 1e-3] from \tableDataDonorMT;

            \addplot[scenario01Dir style, dashed]
            table[x=time_percentage, y expr = \thisrow{LB_False} * 1e-3] from \tableDataNodeUE;

            \addplot[scenario322Dir style, dashed]
            table[x=time_percentage, y expr = \thisrow{LB_True} * 1e-3] from \tableDataNodeUE;

        \end{groupplot}
   \end{tikzpicture}
	\caption{Number of buffered bits in the \ac{mIAB} \ac{MT} to be transmitted in the \ac{UL} backhaul (dotted), \ac{IAB} donors to \ac{mIAB} links (solid) and in the \ac{mIAB} to passengers links (dashed).}
	\label{FIG:buffer_donor_to_mt_ue}
\end{figure}
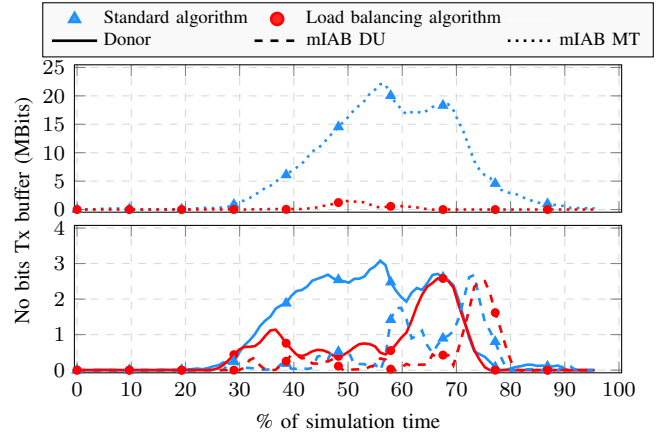

In~\FigRef{FIG:buffer_donor_to_mt_ue}, we present the time evolution of the amount of buffered bits in different nodes in the system: bits in \ac{mIAB} \ac{MT} (dotted curves), \ac{IAB} donors (solid curves) and \ac{mIAB} \ac{DU} (dashed curves). %
The curves for standard solution and proposed solutions are represented by blue and red curves, respectively. %

%
%

Focusing firstly in the \ac{UL} backhaul, i.e., \ac{mIAB} \ac{MT}, we can see that our proposed solution results in a significantly lower accumulation of buffered bits, especially halfway through the simulation, when the bus is closer to the central \ac{IAB} donor, as expected, since the connection between the \ac{mIAB} node and the overloaded \ac{IAB} donor is prevented. %
When the \ac{DL} access and backhaul at \ac{IAB} donors are regarded, we can also observe a lower bits accumulation for our proposed solution. %
However, the decrease in buffer accumulation with our proposal compared to the standard solution is not at the same level as in \ac{UL} backhaul \ac{mIAB} \ac{MT}. %
The reason is that interference becomes more critical in the \ac{DL} when the bus passes in front of the second block, since the closest \ac{IAB} donor (the one at the central block) providing the highest \ac{RSRP} becomes a major source of interference to the backhaul in \ac{DL}. %
This higher interference leads to lower throughput that causes the buffer clogging. %

\FigRef{FIG:buffer_donor_to_mt_ue} also shows the number of bits at the \ac{DL} access of \ac{mIAB} \ac{DU}. %
Interestingly, the load accumulation, in this case, takes place mainly during the last part of the simulation. %
This happens because after the bus passes through the second block, it finally connects to the rightmost \ac{IAB} donor which has more resources to serve it in the \ac{DL} backhaul. %
Then, the rightmost \ac{IAB} donor starts to send the packets that where previously clogged in the overloaded \ac{IAB} donor (in the central block), which in turn overloads temporarily the \ac{mIAB} node buffer. %
	We can also see that when using~\AlgRef{ALG:algorithm_load_balancing}, the buffer size accumulation is lower at first but becomes higher during 70\% to 80\% of the simulation time. 

%

\subsubsection{\ac{UL} Throughput}
\label{SUBSUBSEC:uplink_throughput}

\begin{figure}[t]
	\centering
	    \begin{tikzpicture}
        \begin{axis}[common plots axis options,
            ylabel=CDF,
            xlabel=Throughput (Mbps),
            legend style={
                at = {(0.5, 1.05)},
                anchor = south,
                legend columns = 2,
            }
            ]
            \pgfplotstableread [col sep=comma] {\plotsDataPath/cdf_throughput_general_ped_pass_UL.csv}\tableData

            \addlegendimage{scenario01Dir style, mark size = 1.5pt, only marks}
            \addlegendentry{Standard algorithm}

            \addlegendimage{scenario322Dir style, mark size = 1.5pt, only marks}
            \addlegendentry{Load balancing algorithm}

            \addlegendimage{common line style, solid}
            \addlegendentry{Passengers}

            \addlegendimage{common line style, dashed}
            \addlegendentry{Pedestrians}

            \addplot[scenario01Dir style, mark repeat=25]
            table[x=BS_1_LB_0_Pass_x, y expr = \thisrow{BS_1_LB_0_Pass_y} * 1e2] from \tableData;

            \addplot[scenario322Dir style, mark repeat=25]
            table[x=BS_1_LB_1_Pass_x, y expr = \thisrow{BS_1_LB_1_Pass_y} * 1e2] from \tableData;

            \addplot[scenario01Dir style, dashed]
            table[x=BS_1_LB_0_Ped_x, y expr = \thisrow{BS_1_LB_0_Ped_y} * 1e2] from \tableData;

            \addplot[scenario322Dir style, dashed]
            table[x=BS_1_LB_1_Ped_x, y expr = \thisrow{BS_1_LB_1_Ped_y} * 1e2] from \tableData;

        \end{axis}
    \end{tikzpicture}
	\caption{\ac{UL} throughput \ac{CDF} for the two studied topology adaptation algorithms.}
	\label{FIG:throughput_all_ues_uplink}
\end{figure}
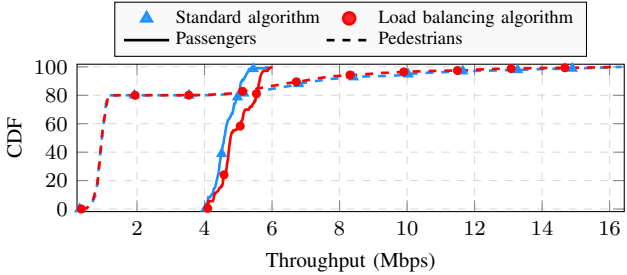

In~\FigRef{FIG:throughput_all_ues_uplink}, we present the \ac{CDF} of passengers (solid curves) and pedestrian (dashed curves) \ac{UL} throughput for the proposed (red curves) and standard solutions (blue curves). %
For passengers, the proposed solution shows throughput gains over the standard solution. %
At the 90\textsuperscript{th} percentile, the throughput for the proposed and standard solutions are of \SI{5.63}{Mbps} and \SI{5.11}{Mbps}, respectively, which leads to a performance gain of approximately 10\%. %
For the 10\textsuperscript{th} percentile, the performance gain is more modest (2.3\%). %
The gains were more prominent in the higher percentiles due to, e.g., the random nature of the pedestrian positioning, which may cause them to become closer to the \ac{mIAB} node causing more interference in some iterations of the simulation. %


For pedestrians, note that our proposal leads to a slight throughput degradation in the highest percentiles. %
The reason for this is that the proposed solution connects the \ac{mIAB} node with the less overloaded \ac{IAB} donors for a longer time. %
Thus, the pedestrians from these \ac{IAB} donors, which have the highest throughputs (highest percentiles in the plot), have to share radio resources with the \ac{mIAB} node leading to a throughput degradation. %
For the same reason, we would expect a throughput improvement for pedestrians at the central block (lower percentiles in the plot) with the proposed solution compared to the standard one. %
However, as there are too many pedestrians to share the system resources in this block, the throughput improvement is not noticeable. %

\subsubsection{\ac{DL} throughput}
\label{SUBSUBSEC:downlink_throughput}

\begin{figure}[t]
	\centering
	    \begin{tikzpicture}
        \begin{axis}[common plots axis options,
            ylabel=CDF,
            xlabel=Throughput (Mbps),
            legend style={
                at = {(0.5, 1.05)},
                anchor = south,
                legend columns = 2,
            }
            ]
            \pgfplotstableread [col sep=comma] {\plotsDataPath/cdf_throughput_general_ped_pass_DL.csv}\tableData

            \addlegendimage{scenario01Dir style, mark size = 1.5pt, only marks}
            \addlegendentry{Standard algorithm}

            \addlegendimage{scenario322Dir style, mark size = 1.5pt, only marks}
            \addlegendentry{Load balancing algorithm}

            \addlegendimage{common line style, solid}
            \addlegendentry{Passengers}

            \addlegendimage{common line style, dashed}
            \addlegendentry{Pedestrians}

            \addplot[scenario01Dir style, mark repeat=25]
            table[x=BS_1_LB_0_Pass_x, y expr = \thisrow{BS_1_LB_0_Pass_y} * 1e2] from \tableData;

            \addplot[scenario322Dir style, mark repeat=25]
            table[x=BS_1_LB_1_Pass_x, y expr = \thisrow{BS_1_LB_1_Pass_y} * 1e2] from \tableData;

            \addplot[scenario01Dir style, dashed]
            table[x=BS_1_LB_0_Ped_x, y expr = \thisrow{BS_1_LB_0_Ped_y} * 1e2] from \tableData;

            \addplot[scenario322Dir style, dashed]
            table[x=BS_1_LB_1_Ped_x, y expr = \thisrow{BS_1_LB_1_Ped_y} * 1e2] from \tableData;

        \end{axis}
    \end{tikzpicture}
	\caption{\ac{DL} throughput \ac{CDF} for the two studied topology adaptation algorithms.}
	\label{FIG:throughput_all_ues_downlink}
\end{figure}
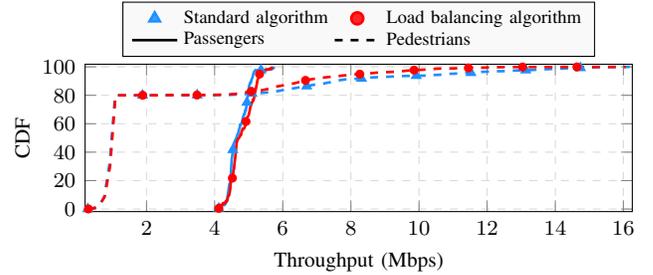

In~\FigRef{FIG:throughput_all_ues_downlink}, we present the \ac{CDF} of passengers (solid curves) and pedestrian (dashed curves) \ac{DL} throughput for the proposed (red curves) and standard solutions (blue curves). %
We can see that there was a general improvement in \ac{QoS} for passengers, but not as expressive as for the \ac{UL} scenario. %
While in the 90\textsuperscript{th} percentile the proposed solution achieves a performance gain of 1.1\%, no performance gain can be seen in the 10\textsuperscript{th} percentile. %
These lower gains in \ac{DL} throughput for passengers can be explained by the increased backhaul interference the \ac{mIAB} node experiences, particularly when it passes in front of the overloaded \ac{IAB} donor (as previously explained). %


For pedestrians, the performance of the proposed and standard solutions are very similar, except at the higher percentiles where \added{our solution presents} a performance loss of 13\% at the 90\textsuperscript{th} percentile. %
The reason is the same as stated for the \ac{UL} performance in~\FigRef{FIG:throughput_all_ues_uplink}: with the proposed solution, the less loaded \ac{IAB} donors have to share their resources for a longer time with the \ac{mIAB} node which leads to performance degradation for the pedestrians with higher throughputs. %
Despite this performance loss, the throughput of \SI{6.67}{Mbps} achieved at the 90\textsuperscript{th} percentile for the proposed solution is still considered a good \ac{QoS} for mobile applications. %

\section{Conclusions}
\label{SEC:conclusions}

This work investigated and compared a standard \ac{RSRP}-only based \ac{TA} algorithm with a proposed solution that additionally takes into account the load in the network nodes. %
Based on our analyses, the \added{proposed} solution significantly reduced the amount of buffered bits, outperforming the standard solution in terms of throughput for passengers. %
Our proposed solution negatively affected the performance of pedestrian \acp{UE}, but mostly the ones with over provision of \ac{QoS}. %
Therefore, the proposed solution effectively enhances the overall \ac{QoS} for passengers without significantly compromising the performance of pedestrians. %

\printbibliography{}
\endrefsection{}
\acbarrier{}

\ifCLASSOPTIONcaptionsoff
	\newpage
\fi

\end{document}